\documentclass[pre,twocolumn,superscriptaddress,aps,notitlepage,10pt]{revtex4-1}
\usepackage{natbib}
\usepackage{times}
\usepackage{amssymb}
\usepackage{amsmath}
\usepackage{bm}
\usepackage{color}
\usepackage[colorlinks=true,urlcolor=blue,breaklinks,citecolor=blue]{hyperref}
\usepackage{graphicx}
\usepackage{braket}
\usepackage{subfigure}
\usepackage{xcolor}

\usepackage[utf8]{inputenc}
\usepackage{hyperref}

\definecolor{fuchsia}{rgb}{1.0, 0.0, 1.0}

\begin{document}

\title{Exploring Quantum Perceptron and Quantum Neural Network structures with a teacher-student scheme} 

\author{Aikaterini (Katerina) Gratsea}
\affiliation{ICFO-Institut  de  Ciencies  Fotoniques,  The  Barcelona  Institute  of Science  and  Technology, 08860  Castelldefels  (Barcelona),  Spain}
\email[]{gratsea.katerina@gmail.com}
\author{Patrick Huembeli}
\affiliation{Institute of Physics, \'Ecole Polytechnique F\'ed\'erale de Lausanne (EPFL), CH-1015 Lausanne, Switzerland}

\begin{abstract}
 \textbf{Abstract} Near-term quantum devices can be used to build quantum machine learning models, such as quantum kernel methods and quantum neural networks (QNN) to perform classification tasks. There have been many proposals on how to use variational quantum circuits as quantum perceptrons or as QNNs. The aim of this work is to introduce a teacher-student scheme that could systematically compare any QNN architectures and evaluate their relative expressive power. Specifically, the teacher model generates the datasets mapping random inputs to outputs which then have to be learned by the student models. This way, we avoid training on arbitrary data sets and allow to compare the learning capacity of different models directly via the loss, the prediction map, the accuracy and the relative entropy between the prediction maps. Here, we focus particularly on a quantum perceptron model inspired by the recent work of Tacchino et. al.~\cite{Tacchino1} and compare it to the data re-uploading scheme that was originally introduced by Pérez-Salinas et. al.~\cite{data_re-uploading}. We discuss alterations of the perceptron model and the formation of deep QNN to better understand the role of hidden units and the non-linearities in these architectures. 
\end{abstract}
\maketitle

\section{Introduction}
Machine learning (ML) combined with quantum information processing gave rise to quantum machine learning (QML), which entails ML tasks performed (at least partially) on quantum computers. Recent results suggest possible advantages for generative QML applications compared to their classical analogs \cite{sweke2020quantum}, but these models are unlikely to run on current quantum devices since they would require a fault tolerant universal quantum computer. In theory, to build a specific QML architectures on a quantum device, e.g. a quantum neural network (QNN), one could always make the classical logic of a neural network (NN) reversible and implement it via unitaries on a quantum computer~\cite{Wan_2017}. This approach would require many qubits and error correction which is unrealistic for current quantum devices either. Furthermore, it is not clear if such an approach would yield any quantum advantage. In recent years, with the advent of Noisy Intermediate-Scale Quantum (NISQ) devices \cite{Preskill_2018} near-term QML applications, such as variational quantum circuits (VQCs) have attracted increasing attention. VQCs are strong candidates for many general classical and quantum optimization applications \cite{qaoa, VQE1, VQE2}, but also for the contruction of quantum perceptrons or neurons \cite{Tacchino1, nonlinearneuron, Wan_2017}. Following the classical analog of NN, QNNs can be built by casting together several of these quantum perceptrons~\cite{Tacchino2, Erik} to build deep structures with hidden layers. In general though, the role of these hidden units and non-linearities in QNNs that are built from VQCs is not as straightforward as for their classical counter part.

Classical deep NNs with many hidden layers of neurons have been successfully applied to different areas, such as computer vision \cite{cvpr} and reinforcement learning \cite{vinyalsGrandmasterLevelStarCraft2019}.
The reason why specifically deep neural networks perform so well on many different tasks is still an active research field in ML~\cite{poggio2017deep}.

In the quantum regime though, it is an open question whether the deep structure of the classical NNs could be reproduced with VQCs. For example, there are proposals for quantum learners that function more in the sense of classical kernel methods \cite{data_re-uploading, huang2020power,schuld2021quantum} and do not allow to build deep structures. Classical kernel methods can be regarded as shallow architectures \cite{towardAI} since they employ non-linear maps on the data inputs. They perform linear regression and the actual optimization, therefore, is convex. The same idea can be applied to quantum kernels, where the non-linear map is provided by the embedding of the data on a quantum circuit and its measurements, the actual optimization though, is again done with linear regression over these kernels. 
In \cite{data_re-uploading}, they also suggest to parameterize the quantum kernel function $\phi(x) = \phi_{\bm{w}}(x)$ with a set of parameters $\bm{w}$ which are part of the learning process, which makes the optimization non-convex but it can still be regarded as shallow.

The aim of this works is to compare the relative expressive power of two conceptually different QML architectures. The first one is the quantum perceptron (QP) as described in \cite{Tacchino1}, which can be cast together with other quantum perceptrons to build deep QNN structures \cite{Tacchino2}. The non-linear activations are introduced with dissipative units where ancillary qubits form the output layer, while the qubits from the input layers are discarded \cite{sharma2020trainability}. The second one is re-uploading architecture (RU) \cite{data_re-uploading} which introduces the non-linear behaviour by repeatedly re-uploading the input data after each trainable unitary. We compare these two architectures to better understand the dissipative nature of the deferred measurement and the re-uploading of the data. To have a fair comparison, we use the same data encoding architecture in both models and to avoid the effect of data selection we deploy a so-called teacher-student scheme, where each architecture will play once the role of the teacher and once the student. Even more, we use different realizations of the teachers and obtain the average performances for the students. This scheme offers a systematic comparison which aims to understand the role of the hidden units and non-linearities in quantum models.

%Both models are dissipative, but the main contribution of the non-linearities for the model in \cite{data_re-uploading} comes from the re-uploading of the data. 
%Therefore, throughout this work we will refer to the former models as QP and the latter models introduced in \cite{data_re-uploading} as the RU.
The paper is structured as follows. Section~\ref{Section4} introduces the quantum architectures that we use throughout this work. Section~\ref{Section5} discuss the ``teacher-student scheme" which gives a systematic comparison between quantum models. Section~\ref{Section6} compares the QP and RU along with some alterations of these models to better understand the role of the hidden units and the non-linearities. Finally, section \ref{Section7} contains the conclusions and outlook. 

\section{Non-linearities in NISQ QNNs}
\label{Section3}
Classical hidden units play a crucial role in the performance and success of neural networks. For example, in Boltzmann machines and feed-forward neural networks, the hidden nodes are feature extractors that capture correlations between the visible nodes. But, while in ffNN the information is propagated through a layer-wise structure of hidden nodes \cite{YOUNES1996}, BMs give an extra perspective on hidden nodes. They extend the sampling space to store conflicting patterns, as happens with the XOR problem in Hopfield networks \cite{wittek2014quantum}. While adding more hidden layers to build a deep NN does not have a straight-forward implementation on near-term quantum devices, extending the sampling space has a clear quantum analog already used in the Born machine \cite{Born_machine} or in dissipative QNNs \cite{sharma2020trainability}.

To build deep structures in QNNs, non-linearities will play a crucial role, as in their classical counterpart. The question that often arises is how to achieve the non-linearities needed in quantum circuits when all the operations are unitary. Just because we have a linear representation of a gate, does not necessarily imply that the gate or the function it computes is linear. A classical logical gate or circuit computing the function $f$ by definition is linear if  $f(x \oplus y) = f(x) \oplus f(y)$, where $\oplus$ is the bit wise XOR operation \cite{XOR-def}. One can, therefore, easily show that for example the Toffoli gate is not linear if applied on binary input strings. Apart from entangling gates, the measurement \cite{Tacchino1, similarTacchino, schuld2020circuit} and the data encoding can introduce non-linearities on NISQ devices. Specifically, the data encoding can introduce non-linearities if the gates are non-linear functions of the data input $x$. For example, if one rotation of the data is performed, the transformation could be linear (for example, with the reflection matrix) and the relative distance of the data is not altered. Instead, if the unitaries perform at least two rotations over different rotation axes the gates are a non-linear function of the data input $x$ \cite{UHUH,data_re-uploading, perezsalinas2021determining}. This ensures the non-linearity of the feature map $x \longrightarrow \psi{(x)}$ in the Hilbert space of the quantum system \cite{schuld2021quantum}.

\section{Quantum models}
\label{Section4}
Exactly like their classical counterpart, quantum perceptrons could either be simple quantum learning machines or the building blocks of quantum neural networks. Both quantum and classical perceptrons are functions that map a set of inputs $\bm{x^{k}}$ to the desired output $y$ by adjusting a set of weights $\bm{w}$. In this work, we explore the notion of non-linearity in quantum neurons along with the formation of deep structures. 

In the quantum regime, the realization of the perceptron and deep QNN is still an active research field and different models have been proposed \cite{Tacchino1, nonlinearneuron}. Different works define quantum perceptron and QNN models as variational quantum circuits (VQC) \cite{Tacchino3, data_re-uploading, sharma2020trainability, congQuantumConvolutionalNeural2019}. Following this approach, for a given data set $\mathcal{D} = \{(\bm{x}^k, y^k) \} $, a quantum circuit $U(\bm{x}^k, \bm{w})$ is parametrized for every input data point $\bm{x}^k$ with label $y^k$ and with trainable parameters $\bm{w}$. The initial state is $\ket{0}^{\otimes N}$, where $N$ is the number of qubits. We denote $\ket{\psi^k} = U(\bm{x}^k, \bm{w}) \ket{0}^{\otimes N}$ as the output state of the quantum perceptron. 
To do classification with only two labels we measure one qubit in a single direction, e.g. in $Z$ direction and interpret the expectation value $\bra{\psi^k} Z \ket{\psi^k}$ as the label prediction.
To train the quantum model, we define the cost function
\begin{equation}\label{cost_function}
    C = \sum_k{ ( y^k - \bra{\psi^k} Z \ket{\psi^k} )^2 },
\end{equation}
which is minimized during the training.
Here, we choose a two dimensional input data set with $\bm{x}^k = ( x^k_1, x^k_2 )$ and encode the data to the circuit with the gates $U(\bm{x}) = R_x (x_1) \otimes R_x (x_2)$ applied on two distinct qubits, where $R_x(\phi) = \exp( -i\phi \sigma_x/2 )$ is the single qubit $X$ rotation gate. These data encoding gates are depicted in light and dark orange colors respectively in the circuit diagrams (e.g. Figure \ref{fig1}) and we will refer to it as the angle encoding. 
Throughout this work, we use the same encoding for all models to have a fair comparison. The trainable part of the perceptron is realized by parameterized single qubit rotations $ Rot(\phi, \theta, \omega) = R_z(\omega)R_y(\theta)R_z(\phi)$ and controlled-Z gates. At the output of each QP we apply a multi-controlled NOT gate inspired by~\cite{Tacchino1}.  The trainable gates are depicted in blue colors in the figure and we will refer to them as processing gates.

Here we focus on two models, the QP and the RU quantum models. The main advantage of the QP is its simplicity and analogy to the classical perceptron~\cite{Tacchino1}, while the RU has attracted a lot of interest recently and has shown great success~\cite{schuld2020effect, data_re-uploading}.

\subsection{Quantum Perceptron (QP)}
The QP is inspired by the recent work of Tacchino et. al.~\cite{Tacchino1}. In their work, the quantum perceptron closely resembles the binary valued classical perceptron. The input is a $m$-dimensional real vector $\bm{x}$ with binary values ${\pm 1}$ for each element $x_i$. The input quantum state $\ket{\psi^k} = U(\bm{x}^k) \ket{0}^{\otimes N}$ is realized by an encoding gate $U(\bm{x}^k)$ acting on $N$ qubits. The output state of the perceptron is defined as $U(\bm{w}) \ket{\psi^k}$, where $U(\bm{w})$ is a parameterized unitary. A multi-controlled NOT gate activates the ancillary qubit which is measured to obtain the quantum perceptron output. The weights $\bm{w}$ are chosen such that the ancillary qubit is activated with a probability $p(\ket{1}_a) = \vert \sum_i^{m} x_i w_i \vert^2$. One can set a threshold for the expectation value $\braket{Z}$ of the ancilla qubit to obtain a binary output from this probability,. 
To build deep structures from QPs one can, like for their classical analog, use the output of one perceptron as the input of the next~\cite{Tacchino2}. Therefore, for a hidden QP, the input state $\ket{\psi^k}$ comes from the ancillary qubit of the previous perceptron and the hidden perceptron itself consists of a parameterized processing gate $U(\bm{w})$, a (multi)-controlled NOT gate that activates another ancillary system and a deferred measurement. This architecture allows one to build deep structures in a coherent way and the ancillary systems can be measured at the very end of the circuit. The idea is that because of this deferred measurement the required non-linearity is still introduced between the layers.

In their original work~\cite{Tacchino1}, the authors focus on binary valued inputs and weights and later they extended it to a continuous values~\cite{Tacchino4}.
Both methods though need the basis encoding which requires many qubits to encode the data. To avoid this, we use the angle encoding introduced earlier in this section. This will now play the role of the gate $U(\bm{x}^k)$, while the parameterized gates $Rot(\phi, \theta, \omega)$ (in blue in the circuit diagrams) together with the CZ entangling gates realize the $U(\bm{w})$ gates for the data processing. This allows for continuous inputs and weights. The perceptron model for the simple case of a two dimensional input is illustrated in Fig.~\ref{fig1}a. 
\subsection{Re-Uploading Architecture (RU)}
The RU is inspired by the recent work of Perez-Salinas et. al. \cite{data_re-uploading}. In their work, they propose a quantum version of a ffNN with only one hidden layer.
To emulate the behaviour of a ffNN, they repeatedly apply data encoding unitaries $U(\bm{x}^k)$ parameterized with the input data $\bm{x}^k$ to the circuit. 
Each of these data dependent unitaries is followed by a parameterized unitary $U(\bm{w})$ with trainable parameters $\bm{w}$. Because of the repeated application of $L_i = U(\bm{w}) U(\bm{x})$ to the circuit, this scheme is called data re-uploading. A single application of $L_i$ is often referred as a re-uploading layer and their number determine the architecture of the re-uploading circuits.

In this work, we use the angle encoding as defined earlier and we parameterize the processing gates with $R(\phi, \theta, \omega)$ and CZ gates. To simplify the measurement procedure and make the two perceptron models more comparable, we also use a multi-controlled NOT gate to activate an ancilla qubit exactly like in QP. We also use the same cost function from Eq. \eqref{cost_function}. 

The notion of depth is not as clearly defined for the re-uploading scheme as it is for the QP. It is known that more layers of re-uploading lead to better expressivity of the model~\cite{schuld2020effect}, but there is no clear classical analog to this. Therefore, we will compare the RU with an increasing number of re-uploading layers $L_i$ to QNN models with an increasing number of QPs.
In the next section, we elaborate the scheme that we use to compare the aforementioned architectures. 
\begin{figure}
\label{fig1}
\includegraphics[scale=1.0]{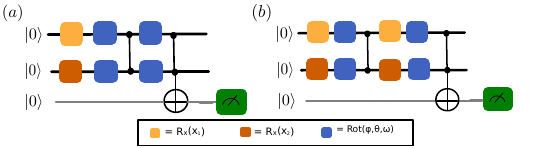}
\caption{ The quantum circuits that represent the quantum perceptron (QP) (a) and re-uploading quantum model (RU) (b).}
\label{fig1} 
\end{figure}
\section{Teacher-student scheme}
\label{Section5}
Here, we introduce the teacher-student scheme that aims to to benchmark different realizations of QP and QNN against each other. We systematically compare the two aforementioned architectures (QP and RU see Fig.\ref{fig1}), but the scheme could be used for any circuits. The main idea is that one architecture (for example the QP) will play the role of the teacher and generate the labels that will be used to train the student (for example the re-uploading quantum model). Thus, we avoid to generate artificial data sets, such as e.g the circle data set in \cite{data_re-uploading}, that could possibly favour one of the architectures. With the teacher-student scheme, we directly see the data structures that each architecture can generate and how well other architectures could learn them.
\subsection{Notion of teacher}
The teacher generates the labels for a fixed set of inputs $\mathcal{D}= \{\bm{x}^k \}$ with 2 dimensional input vectors $\bm{x} = \{x_1, x_2 \}$ on a grid $x_i \in [-\pi, \pi]$. For a fixed teacher architecture we choose several random initializations for the parameters $\bm{w}$ of the processing gates (blue squares in the figures). We use the measurement outcomes of the ancilla qubit as the model predictions $y^k$ of the input data. This way we can generate several data sets for different random initializations. The predictions have continuous values $y^k \in [-1, 1]$ given by the outcome of the measurement $ \bra{\psi^k} Z \ket{\psi^k} $, but we also generate binary valued labels by choosing $ y^k_{\text{binary}} = \text{sign} (y^k)$. The teachers with binary labels focus more on the basic characteristics of the data structures, while the ones with continuous labels also care for the details. We can visualize the data structures with $\it{prediction}$  $\it{maps}$, which are the density plots of the model predictions and labels $y^k$ for the input data $\bm{x}^k$.
\subsection{Notion of student}
We train the students with the labeled data generated by their teachers to learn those data structures. It is not obvious how to define a good/bad student, since different tools can be used to characterize their performance. The $\it{prediction}$  $\it{maps}$ of the students are best for visualizing the similarity of the student's and teacher's predictions of the label $y^k$ to gain qualitative results. For a more rigorous quantitative comparison we compute the $\it{relative}$ $\it{entropy}$ between the student's and teacher's outputs $y^k$. Specifically, we use the information divergence (Kullback–Leibler divergence or relative entropy) which defines a distinguishable measure between two probability distributions $P$ and $Q$~\cite{geometry_of_quantum_states}: 
\begin{align}
S(P \| Q)=\sum_{i=1}^{N} p_{i} \ln \dfrac{p_{i}}{q_{i}}.
\end{align}
When the two distributions are similar, the value of the relative entropy is close to zero. To interpret the predicted labels $y^k$ as probabilities, we offset and re-normalize them ($y^k>0$, $\sum_{\bm{x^k}\in \mathcal{D}} y^k = 1 $). Then, to compare two prediction maps, the information divergence is calculated by summing over the whole input space. Here, we are interested in the average relative entropy of all teacher-student pairs.  Another qualitative metric is the $\it{loss}~\it{function}$ which determines the success of the training. When the student is trained with the binary valued labels the percentage of the correctly predicted labels can be computed. We refer to this as the $\it{accuracy}~\it{score}$ which gives an overall performance of the student. To identify a good/bad student all these tools should be taken into account.
\section{Results}
\label{Section6}

\begin{figure}
\includegraphics[scale=0.4]{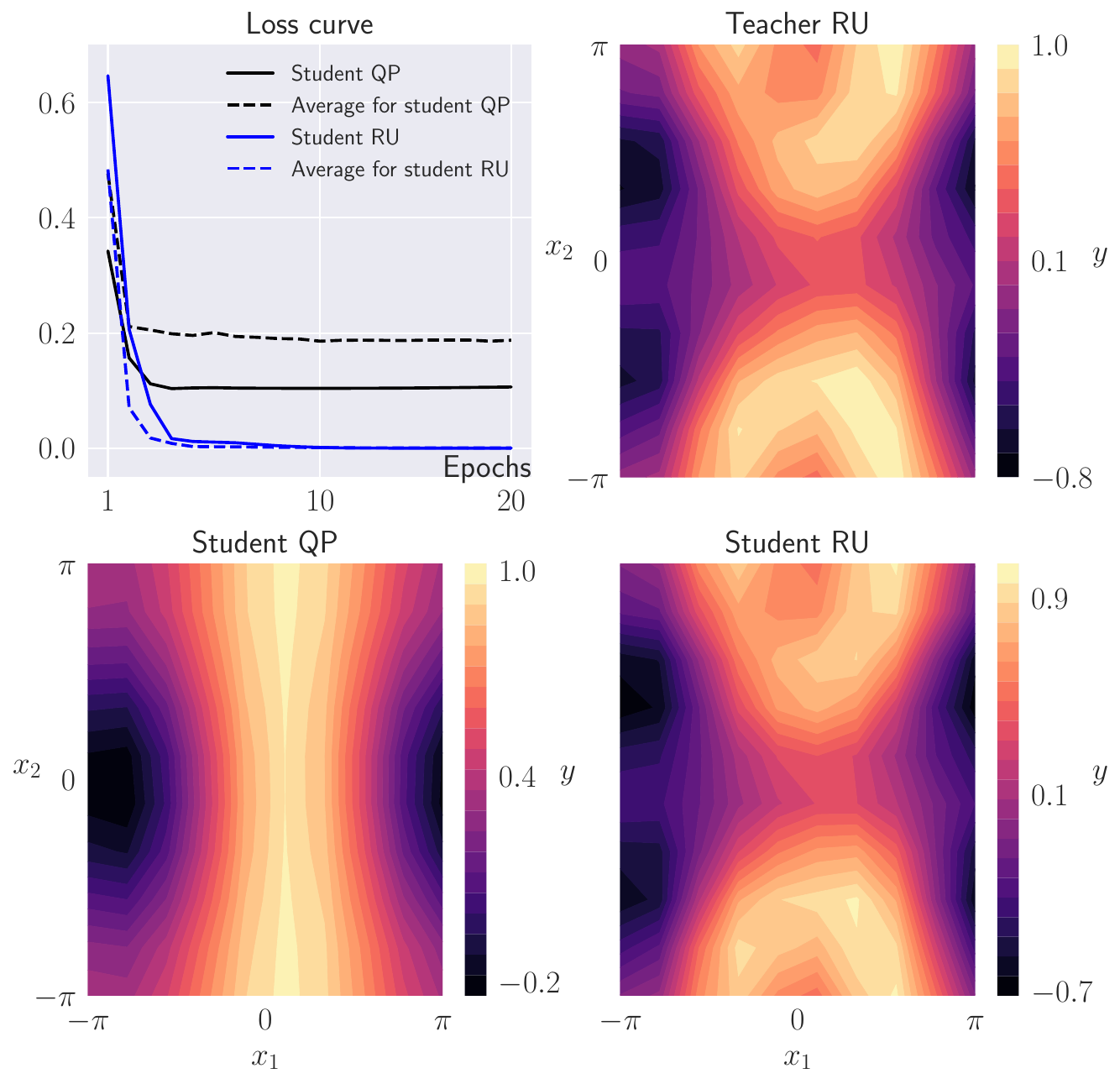}
\caption{Teacher-student training for the teacher RU (Fig.~\ref{fig1}b) and 2 students (QP and RU in Fig.~\ref{fig1} a and b, respectively). The prediction maps show one particular example of the training with the corresponding losses (solid loss curves). The average loss (dashed lines) show the average over all 10 random initializations. The teacher-student training for the teacher QP can be found in the appendix Fig.\ref{fig7A-all}. }
\label{fig2}
\end{figure}

\subsection{Toy model}\label{sec_Toy_model}
Here, we will elaborate the notion of the teacher-student scheme with an illustrative example. We use the RU architecture as a teacher  (Fig.~\ref{fig1}b) and both, QP and RU, as students (Fig.~\ref{fig1}a and b). A very characteristic example for the prediction maps is shown in Fig.~\ref{fig2}, where the Student - QP does not learn the inner structure (circles) at the left and right sides of the prediction map of the teacher. The Student - RU reproduces almost perfectly the teacher as expected, since they have the same circuit architecture. But the Student RU finds a different set of parameters compared to the one of the Teacher RU. We present the parameters of the processing gates for both of them in the Appendix (\ref{parameters}). The solid lines in the ``Loss curve" in Fig.\ref{fig2} show the loss of the two students that correspond to these particular prediction maps, where Student - Re-uploading achieves a much lower loss. If we use the binary labels of the teacher and train again the students we reach the accuracy score approximately equal to $0.9$ compared to $0.8$ for the QP. Both students have a high accuracy score, since the topology of the prediction maps of the students and the teacher with binary labels are very similar (see Appendix \ref{add_plots}, Fig.~\ref{fig2A-all}).

We generated 10 different data sets by randomly initializing the teacher's parameters as explained earlier to obtain the average performance of the student. Both students are trained on those data sets and we characterize their performance with the tools mentioned earlier. The dashed lines in Fig.~\ref{fig2} show the average of the loss function over all the initializations of the teacher for both students. The Student - RU shows good convergence (blue dashed line) for all data sets generated by the teacher, since the teacher and student have the same architecture in this Toy example. The average loss for the Student - QP (black dashed line) is larger than the single example shown in the prediction map (black solid line), which suggests some of the teacher's data sets can be learned more accurately by the student than others. These results are also supported by the calculation of the average relative entropy over all prediction maps which is equal to $0.247$ and $0.001$ for the Student - QP and the Student - RU, respectively. It is worth noting here that the loss captures better the global differences in the prediction (i.e. a general off-set of the whole prediction maps), while the relative entropy captures local differences in the maps (i.e. the local minima that do not appear in one of the maps). For more details, we provide the explicit
code used for this toy model in~\cite{Git-repo}.

In order to have a complete comparison of the two models, we now use the QP architecture as the teacher and train again for both students (see Fig.~\ref{fig2-appendix} in Appendix). The average loss for both students converges to similar low values (see Fig.~\ref{fig2-appendix} in Appendix) and we have a high accuracy score $~0.9$ of both students. The average relative entropy is approximately equal to zero for both students ( $0.0002$ for the QP and $0.0013$ for the RU). These results suggest that the QP generates simpler data structures and both architectures can learn them. 

In conclusion, the RU architecture can generate and learn more complex output distributions than the QP. It can learn all the data sets generated by the QP with almost zero loss and relative entropy. On the contrary, the QP is not able to learn the data sets provided by the RU. Even for this toy example, all four metrics defined earlier were taken into account, which reveals the difficulty on how to determine whether a student is good or bad. Therefore, one should take under consideration all the different tools introduced earlier (the prediction maps, the accuracy score, the (average) loss and relative entropy) to characterize the performance of the students.
\subsection{A more complex teacher}
\begin{figure}
\includegraphics[scale=0.45]{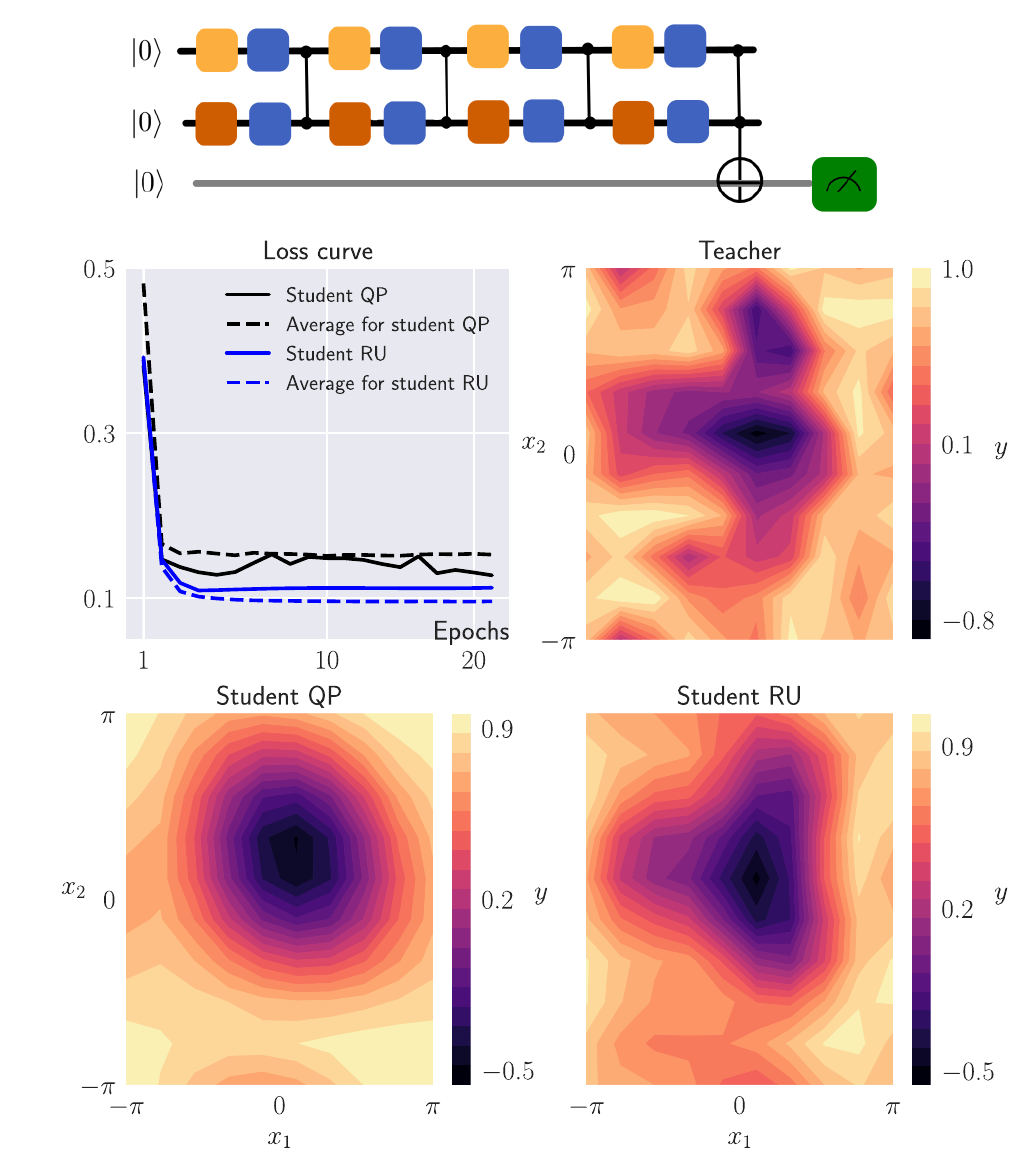}
\caption{The ``deep"  architecture of the teacher in terms of the number of times the data are encoded. The prediction maps of the teacher with the deep architecture and the students: Student - QP and Student - RU as in Fig.~\ref{fig1}. In the upper right corner, we plot the loss curves for the Student - QP (black line), Student - RU (blue line) and the student averages over 10 different realizations of the teacher (Student - QP: black dashed line and Student - RU: blue dashed line).}
\label{fig3} 
\end{figure}

Next, we compare the behaviour of the two models in Fig.~$\ref{fig1}$ (Students) to a ``deep" re-uploading architecture in Fig.~$\ref{fig3}$ (Teacher) with many repetitions of the encoding gates. In Fig.~$\ref{fig3}$, we show a characteristic result for the prediction maps. The Student - RU learns the basic features of the teacher's architecture contrary to the Student - QP which learns a very simplified version of the data structure. As mentioned earlier, the number of encoding unitaries determine the function complexity that can be learned \cite{schuld2020effect}. Therefore, the teacher with four pairs of encoding gates generate a more complex data structure compared to the ones that the student of the QP and RU can learn with one and two pairs of encoding unitaries, respectively. If we train the students on the binary valued outputs, we reach accuracy scores approximately equal to $0.8$ for both students, because the topology of the prediction maps of the students and the teacher with binary labels are very similar (see Appendix \ref{add_plots} Fig.~\ref{fig7A-all}). Therefore, the accuracy score does not allow to determine which student learns the teacher more accurately. But the prediction maps for this characteristic example (see Fig. \ref{fig3}) shows the better performance of the Student - RU. Even more, the loss curves, and specifically the average loss curves, support this claim. As we can see from Fig.~\ref{fig3}, the average loss curve for Student - RU (blue dashed line) over 10 realizations of the teacher is much lower than the average loss curve of Student - QP (black dashed line). Finally, the average relative entropy is equal to $0.15$ and $0.11$ for the Student - QP and the Student - RU, respectively and enhances this statement.
\subsection{Improving the QP}
\begin{figure}
\includegraphics[scale=1.0]{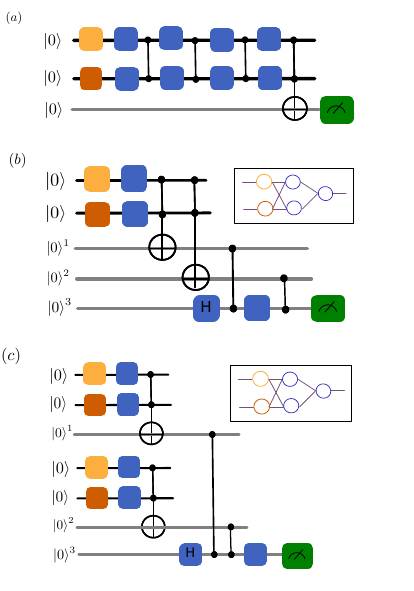}
\caption{The alterations of the Student - QP to an eight gate QP (a), a QNN model (b) and a quantum ffNN as proposed in \cite{Tacchino2}. In the insets, the analogy to classical deep NN is shown for the circuits in (b) and (c). The models in (a) and (b) have the same expressivity with the QP, contrary to the quantum ffNN (c) which performs better as the RU (see Fig.~\ref{fig5}).}
\label{fig4} 
\end{figure}

Here, we present some alterations of the Student - QP to improve its performance and further understand its behaviour. So far, we used four general single qubit rotations and two 2-qubit gates (the multi-controlled NOT included).
Sousa et. al. \cite{sousa2006universal} show that eight single qubit unitaries and four entangling gates realize an arbitrary two qubit gate.  Therefore, we add four more processing gates to the Student - QP each followed by CNOT entangling gates (see figure \ref{fig4}a). We still have a single perceptron with two inputs, but with a more general unitary for processing the data. The performance of the perceptron though did not improve, suggesting that the structure of the circuit is not enough to simulate the data structure of the re-uploading perceptron as a teacher. 

\begin{figure}
\includegraphics[scale=0.38]{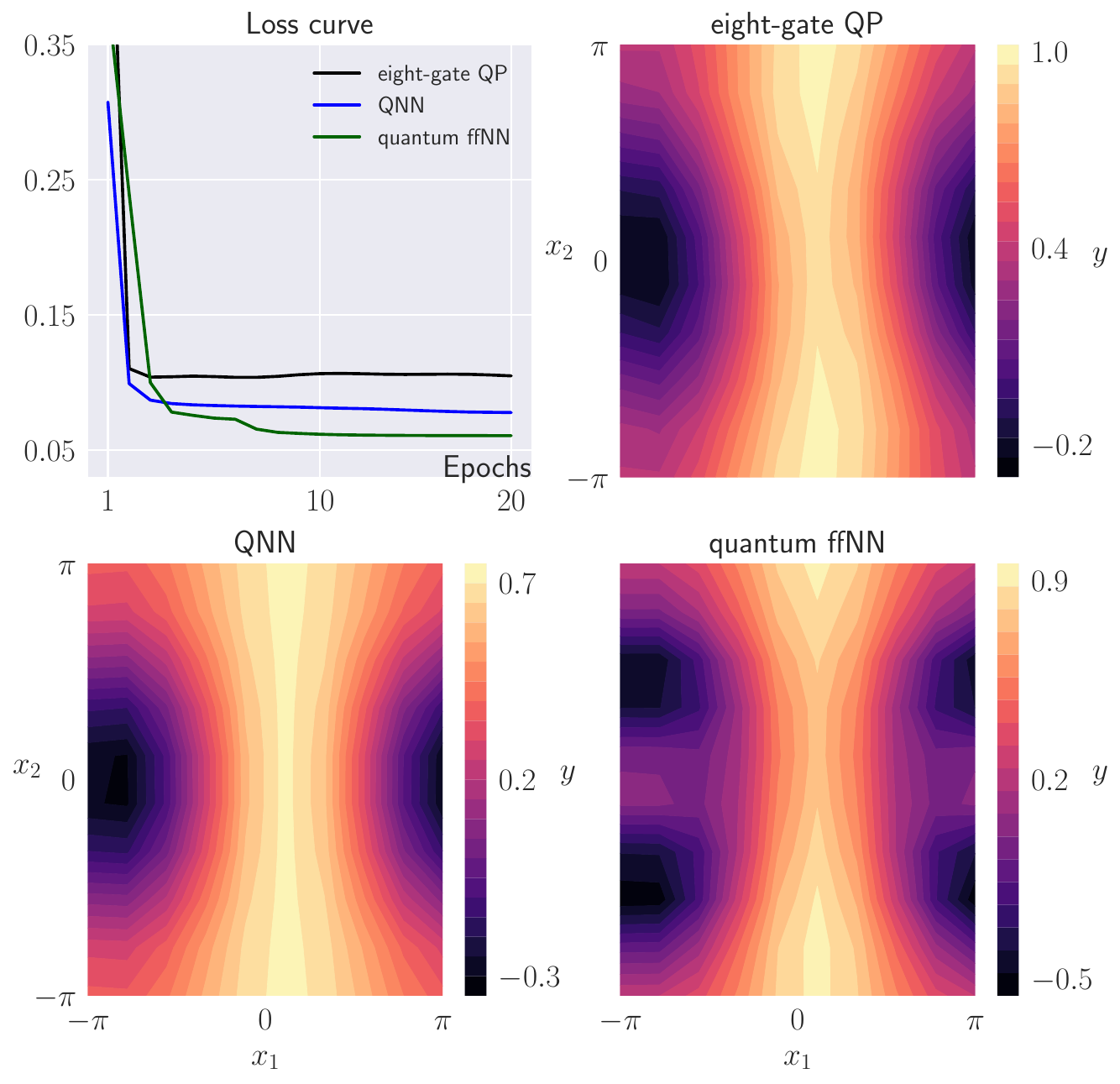}
\caption{The prediction maps of the students in Figures \ref{fig4}a, \ref{fig4}b and \ref{fig4}c and their loss curves.}
\label{fig5}
\end{figure}

Next, we try to increase the complexity of the QP by realizing a deep QNN from single QPs inspired by~\cite{Tacchino2}. The structure is shown in Fig.  \ref{fig4}b with the classical analog in the inset which can be understood in the following way: The input data (light and dark orange circles in the inset) are introduced with the encoding unitaries (light and dark orange squares). The processing happens with the blue unitaries which correspond to the blue circles in the inset. The output of each blue circle is realized with the CNOT gate acting on both ancilla qubits 1 and 2. These outputs are then introduced in the last ancilla qubit 3 with the CZ gates. The performance of the circuit did not improve though (see Fig.~\ref{fig5}), which means that the deferred measurement on the second ancilla is not enough to improve the performance of the architecture. This shows that the formation of coherent deep QNN structures does not follow the classical analog of a classical deep NN, i.e. by staking several perceptrons together.

Another alteration of the QP is shown in Fig.~\ref{fig4}c. There we use the QNN structure that Tacchino et. al. proposed in \cite{Tacchino2}, where the QP is used as a building block to create the QNN, exactly like in the classical case. Specifically, instead of using the output of one QP twice as in Fig.~\ref{fig4}b, we use two QP to form a QNN. This circuit shows improved learning capability and performs quite well as suggested by the convergence to a low value of the loss curve (see Fig.\ref{fig5}). Therefore, comparing the QNN (Fig.~\ref{fig4}c) with the deep QP (Fig.\ref{fig4}b) suggests that the deferred measurement does not provide the necessary depth to the circuit and shows the importance of re-uploading the data several times to increase the performance of the architecture.

Even though the QNN architecture in Fig.~\ref{fig4}c seems quite complex (many qubits and entangling gates), at the end is very similar to the circuit architecture of the RU (Fig.~\ref{fig1}b). Both circuits encode the data twice in a vertical or in a horizontal way, respectively. For the vertical case $2N$ qubits are used for $N$ encodings of the data (see Fig.~\ref{fig4}c), while in the horizontal case the $N$ encodings apply in the same two qubits (see Fig.~\ref{fig1}b). The recent work of Schuld et. al. \cite{schuld2020effect} shows that the complexity of the functions that a circuit learns is determined by the number of times that the encoding gates are applied. They emphasize that there is no difference if the encoding data are introduced in a circuit vertically (as in the QNN Fig.~\ref{fig4}c) or horizontally (as in the RU Fig.~\ref{fig1}b). In that sense, the architectures of these two models are equivalent and the prediction maps enhance this claim. But, as suggested in the work \cite{schuld2020effect}, the trainable circuit and measurement could affect the functions that can be approximated. We already see here the qualitative difference of these two models. For the QNN in Fig.~\ref{fig4}c, the average accuracy score is equal to $0.84$ and the average relative entropy gives $0.098$ which is significantly different from the values of the RU in Fig.~\ref{fig1}b ($0.9$ and $0.0009$, respectively). This suggests that the architecture Fig.~\ref{fig4}c has similar performance to the RU, but as the metrics suggest there are some qualitative differences between them due to the differences at the trainable part of the circuits \cite{schuld2020effect}.

\subsection{Role of the encoding}
Contrary to classical perceptrons, quantum perceptrons can have different encodings, i.e. amplitude, basis and angle encoding~\cite{book_Schuld}. In this work, the encoding of the two-dimensional data $\bm{x} = (x_1, x_2)$ is achieved with the application of $R_x (x_1)$ and $R_x (x_2)$ to two distinct qubits. A good choice of the encoding can immediately lead to a linear separation of the data as suggested in recent works \cite{robust_data_encoding, data-encoding_Schuld}. Therefore, exploring a quantum perceptron model is interconnected with exploring the encoding of the input data. 
Here, we want to compare the aforementioned angle encoding with the encoding $  Rot(x_1, x_2, 0) H$ applied at each qubit. The Hadamard gate $H$ is applied to $\ket{0}$, since otherwise for $Rot(\phi, \theta, \omega) = R_z(\omega)R_y(\theta)R_z(\phi)$, the first gate $R_z(x_1)$ applied to $\ket{0}$ would not contribute anything. 

\begin{figure}
\includegraphics[scale=0.4]{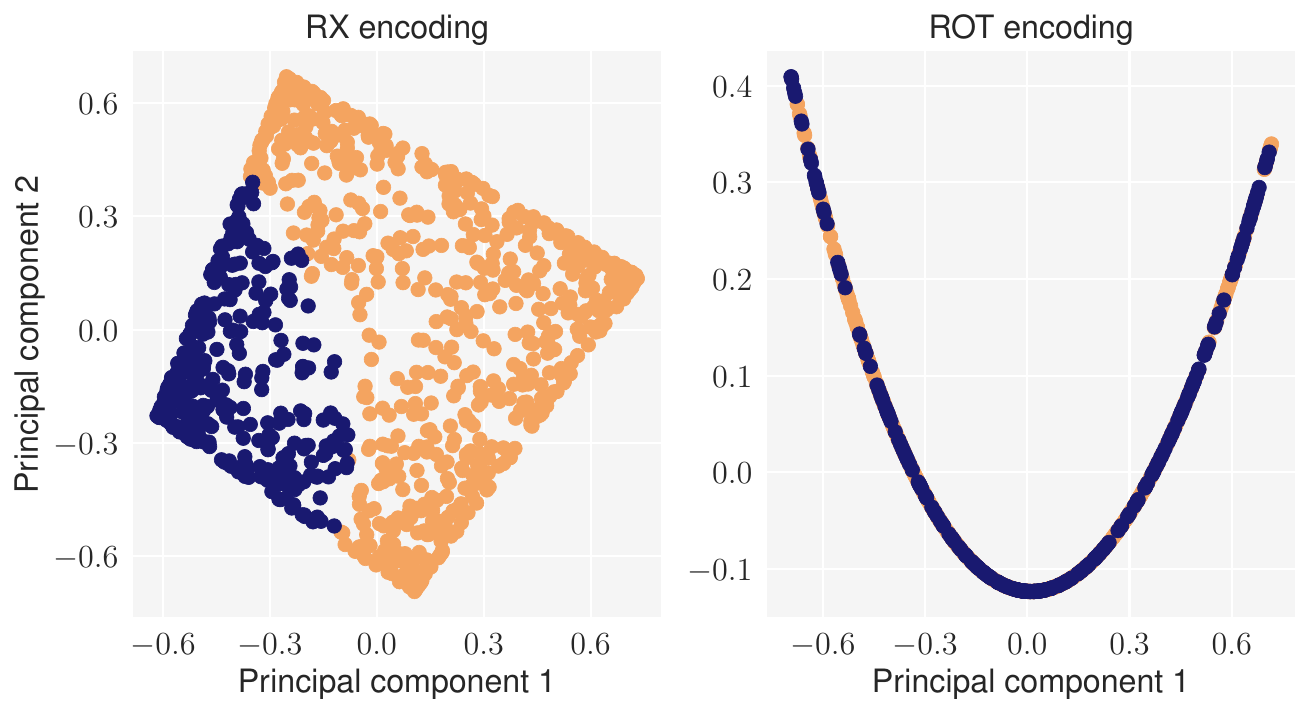}
\caption{PCA on the probability distribution for the $RX$ (left) and $Rot$ (right) encodings. One can see that the circuit data immediately separates for the $RX$ encoding but it does not for the $Rot$.}
\label{PCA} 
\end{figure}

To see how important the encoding is to separate a predefined data set, we apply the encoding for all data $\bm{x}^k$ from a circular data set (see Fig.~\ref{fig11} in Appendix) without the processing unitaries and generate the states $\ket{\psi^k} = U(\bm{x}^k) \ket{0} = c_{00} \ket{00} + c_{01} \ket{01} + c_{10} \ket{10} + c_{11} \ket{11}$ for both encodings. Specifically, we have $\ket{\psi^k}_{RX}$ for $U(\bm{x}^k) = R_x(x_1) \otimes R_x(x_2)$ and $\ket{\psi^k}_{Rot}$ for $U(\bm{x}^k) = Rot(x_1, x_2, 0) \otimes H \otimes Rot(x_1, x_2, 0) H$. We find the probability vectors $P = \left[ p_{00}, p_{01}, p_{10}, p_{11} \right]$ for each of the states $\ket{\psi^k}$, with $p_{i,j} = \vert c_{i,j}\vert^2$.
We are interested in the probability vectors and not the quantum states themselves because for our perceptron models, the probability $p_{11}$ corresponds to the activation probability of the Toffoli gate. Specifically, we encode $500$ input data points $\{x_1, x_2\}$ from the circular data set and create the probability vectors for each one of them for the different encodings. We apply principle component analysis (PCA) on each one of the probability vectors which converts them to two dimensional data sets. Then, we can see whether the data are already separable just after the encoding. In Fig.$\ref{PCA}$, we plot these two dimensional data sets for both encodings. With the RX encoding the data are already separable, contrary to the Rot encoding where the data are along the same curve. 
The data encoded with $RX$, shown in Fig.\ref{PCA}(left), can be separated with a quadratic function as discussed in the work \cite{robust_data_encoding} and a single general parameterized unitary $U(\bm{w})$ that post processes the state $\ket{\psi^k}$ would suffice. The data in \ref{PCA}(right) cannot be separated by a single parameterized processing unitary and would need for example more data re-uploadings.

The aforementioned results stress the importance of the encoding in a quantum circuit~\cite{schuld2020effect} and shows that synthetic data sets can favour certain architectures. If the data are already separated after the encoding gates, the parameterized gates will simply rotate the data along the measurement axis. But, if the encoding fails to separate the data as in the case of $Rot$, the parameterized gates will not be able to separate them on their own. Also, we want to emphasize that the re-normalization of the data affects the success of the training, since the quantum functions are periodic and the input data should lie within the period of the function \cite{Schuld_Kiloran} (see Appendix \ref{input_data} Fig.~\ref{fig5-zoomin}). Finally, it worths mentioning that the encoding of the data labels could also affect the success of the training (see Appendix \ref{Labelling} Fig.~\ref{fig11}).

\section{Conclusions}
\label{Section7}
Inspired by the recent works \cite{data_re-uploading, Tacchino1}, we explored the expressive power of QPs, their formation to QNNs and the RU models implemented on NISQ devices. In order to systematically compare the architectures, we introduced a so-called teacher-student scheme, where the studied models are introduced once as a teacher and once as a student. This way we can avoid to generate synthetic data sets that might give an advantage to certain architectures and it creates a more fair framework for comparing any quantum models. 

Specifically, we showed that the deep structures that can be built with QPs only increase the expressivity of a model if the data are uploaded several times. It is not sufficient to use deferred measurements to generate hidden non-linearities similar to classical NNs if the output of a QP is reused. 
We explored several different ways of how to leverage deferred measurements to generate hidden non-linearities, but the expressive power of QNNs only improved when additional data-uploadings were added (Fig.~\ref{fig4}c). This suggests that the non-linear behaviour induced by a measurement of a single QP cannot be generalized to deep QNNs if the single QPs are cast together in a coherent way (Fig.~\ref{fig4}b). 
Therefore, it is still an open question how to build deep QNNs in a coherent way, where measurements only occur at the end of the computation. Thus, one should not expect a one-to-one mapping of quantum and classical NNs.

These results are in accordance with the recent work \cite{Schuld_Kiloran}, which shows that the number of times that the data are encoded determines the functions that can be approximated. The needed non-linearities in a quantum model can be generated (apart from the measurement) from the encoding gates that are non-linear functions of the input data. Performing PCA on the probability vectors, we showed that given the encoding, the data can already be separated without further processing. Therefore, the performance of a QP is strongly affected by the encoding and the dataset itself. Apart from the encoding, the processing plays an important role as well. The universal approximation capability of different quantum models has been discussed extensively in \cite{data_re-uploading, schuld2020effect, perezsalinas2021determining, perezsalinas2021qubit}, but it does not provide any information about how well the circuit could perform or how many parameters it needs to approximate a  function within a certain error. The calculation of the average relative entropy for Fig.\ref{fig4}c and Fig.\ref{fig1}b showed that the trainable part of the circuit affects the functions that can be approximated. This effect will be further explored in subsequent research.

For future work, it will be of great interest to explore different perceptron models and compare their performance with the teacher-student scheme. Then, the question arises which perceptron model will be the ideal building block of QNN architectures and how quantum perceptrons could be combined to form a deep QNN. Another research direction is to explore other quantum models with no direct analog with classical NN, like the re-uploading model or quantum kernels in general. Finally, it would be of great importance to further explore the role of entanglement and encoding in QPs, in their formation to QNNs and in other quantum models.

\section{Code}
Code to reproduce the results for the toy model~\ref{sec_Toy_model} and explore further settings can be found in the following Github repository: 
\url{https://github.com/KaterinaGratsea/Teacher-student_scheme}.

\section{Acknowledgements}
We would like to thank Mark Fingerhuth, Gorka Munoz Gil and Alexander Dauphin for their useful feedback and critical thinking on the manuscript. The simulations were made with the Pennylane library~\cite{bergholm2018pennylane} and the graphics with the Xournal++ software.
A.G acknowledges support from ERC AdG NOQIA, Spanish Ministry of Economy and Competitiveness (“Severo Ochoa” program for Centres of Excellence in R\&D (CEX2019-000910-S), Plan National FIDEUA PID2019-106901GB-I00/10.13039 / 501100011033, FPI), Fundació Privada Cellex, Fundació Mir-Puig, and from Generalitat de Catalunya (AGAUR Grant No. 2017 SGR 1341, CERCA program, QuantumCAT  \textunderscore U16-011424 , co-funded by ERDF Operational Program of Catalonia 2014-2020), MINECO-EU QUANTERA MAQS (funded by State Research Agency (AEI) PCI2019-111828-2 / 10.13039/501100011033), EU Horizon 2020 FET-OPEN OPTOLogic (Grant No 899794), and the National Science Centre, Poland-Symfonia Grant No. 2016/20/W/ST4/00314. This project has received funding from the European Union's Horizon 2020 research and innovation programme under the Marie Skłodowska-Curie grant agreement No. 847517 and 665884.

\renewcommand{\bibsection}{\section*{References}}

\bibliography{main.bib}

\appendix 
\section{Appendix}\label{Appendix}

\subsection{Parameters} \label{parameters}
The parameters of the four arbitrary single qubit gates $[ \phi_i, \theta_i, \omega_i ]$ of Teacher RU are 
\begin{align*}
[ [2.71, 6.15, 0.42], [3.26, 1.13, 6.10], \\
[0.71, 2.54, 4.63], [4.43, 2.66, 2.18] ],
\end{align*}
while for the Student RU we have
\begin{align*}
[ [-1.30, 0.40, -1.86], [0.60, 1.89, 0.37],\\
[0.17, 2.69, 0.67], [0.49, -0.14, 0.68] ].
\end{align*}
This is one of the many local minima that the training of Student RU could reach.

\subsection{Additional plots}\label{add_plots}
\begin{figure}[h]
\includegraphics[scale=0.4]{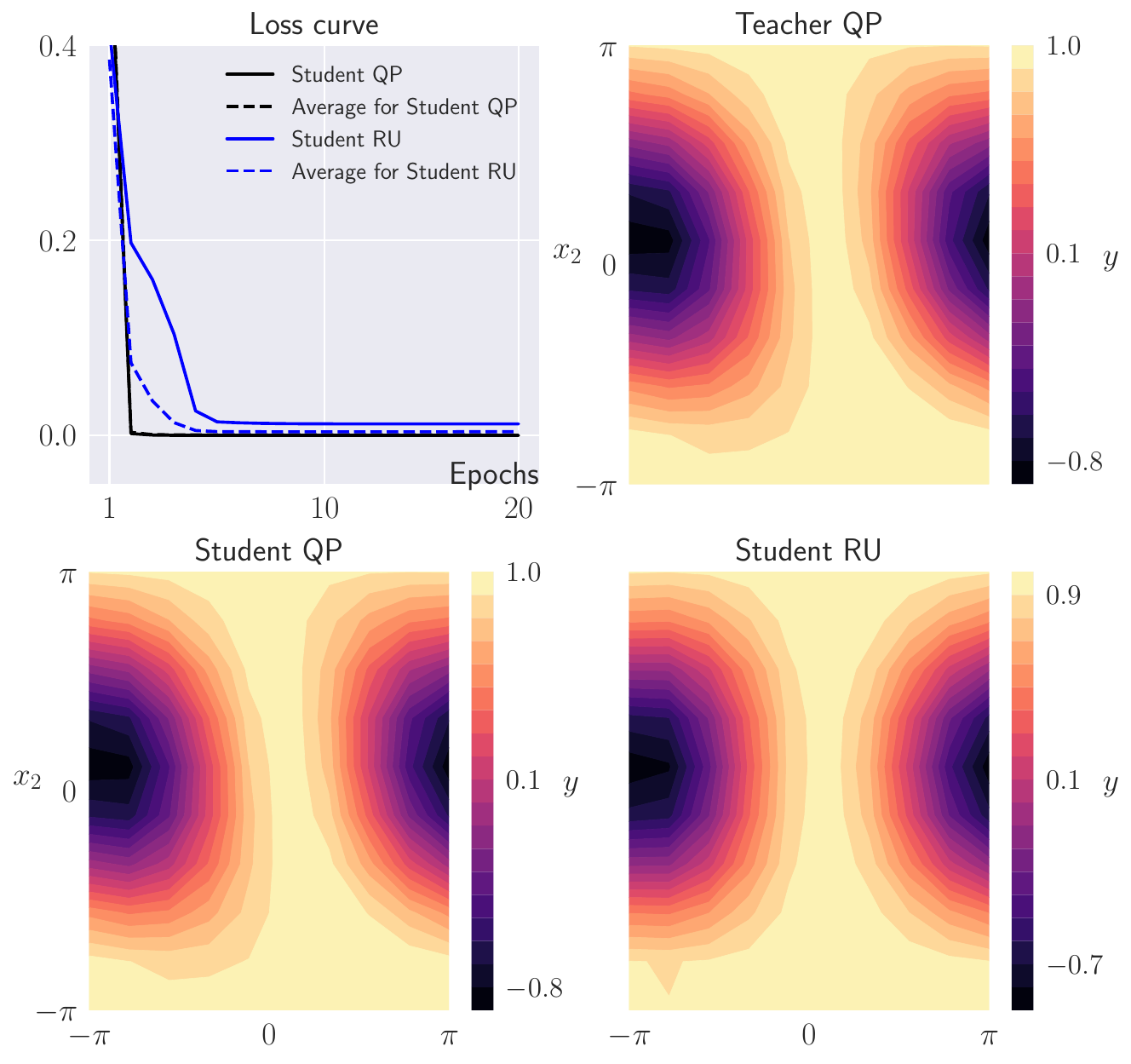}
\caption{Teacher-student training for teacher (Fig.~\ref{fig1}a) and the two students (Fig.~\ref{fig1} a and b). The prediction maps show one particular example of the training with the corresponding losses (solid loss curves). The average loss (dashed lines) show the average over all 10 random initializations.}
\label{fig2-appendix}
\end{figure}

In Fig.~\ref{fig2-appendix}, we show the prediction maps and loss curves for the toy model with the QP (Fig.~\ref{fig1}a) as the teacher and the students (Fig.~\ref{fig1}a and b).

\begin{figure}
\includegraphics[scale=0.38]{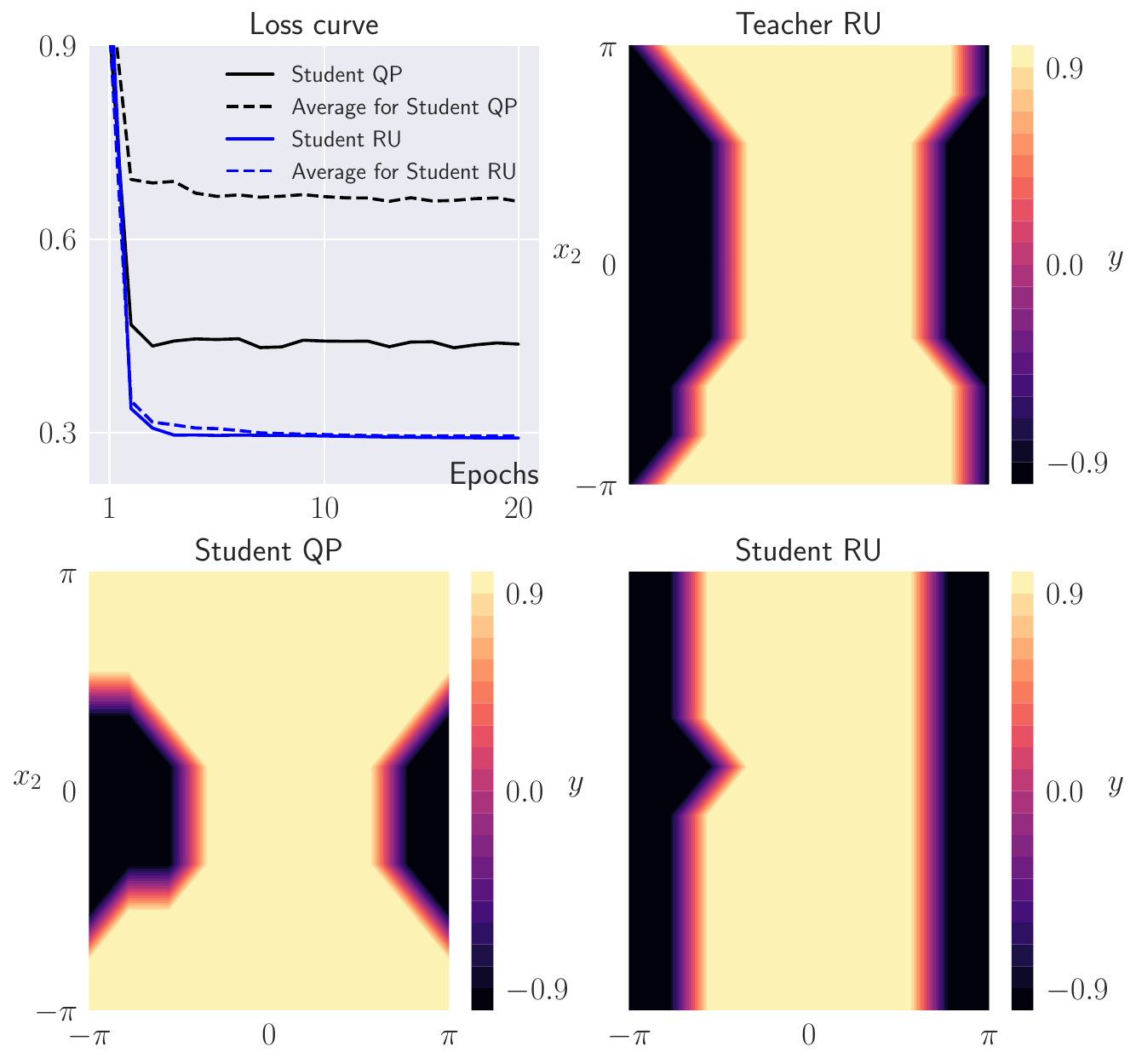}
\caption{The binary prediction maps of the teacher RU and the students: Student - QP and Student - RU as in Fig.~\ref{fig1}. In the upper right corner, we plot the accuracy curves for the QP (Student - QP black line), RU (Student - Re-uploading blue line) and the student averages over 10 different realizations of the teacher (Student - QP black dashed line and Student - Re-uploading blue dashed line).}
\label{fig2A-all}
\end{figure}
\begin{figure}
\includegraphics[scale=0.38]{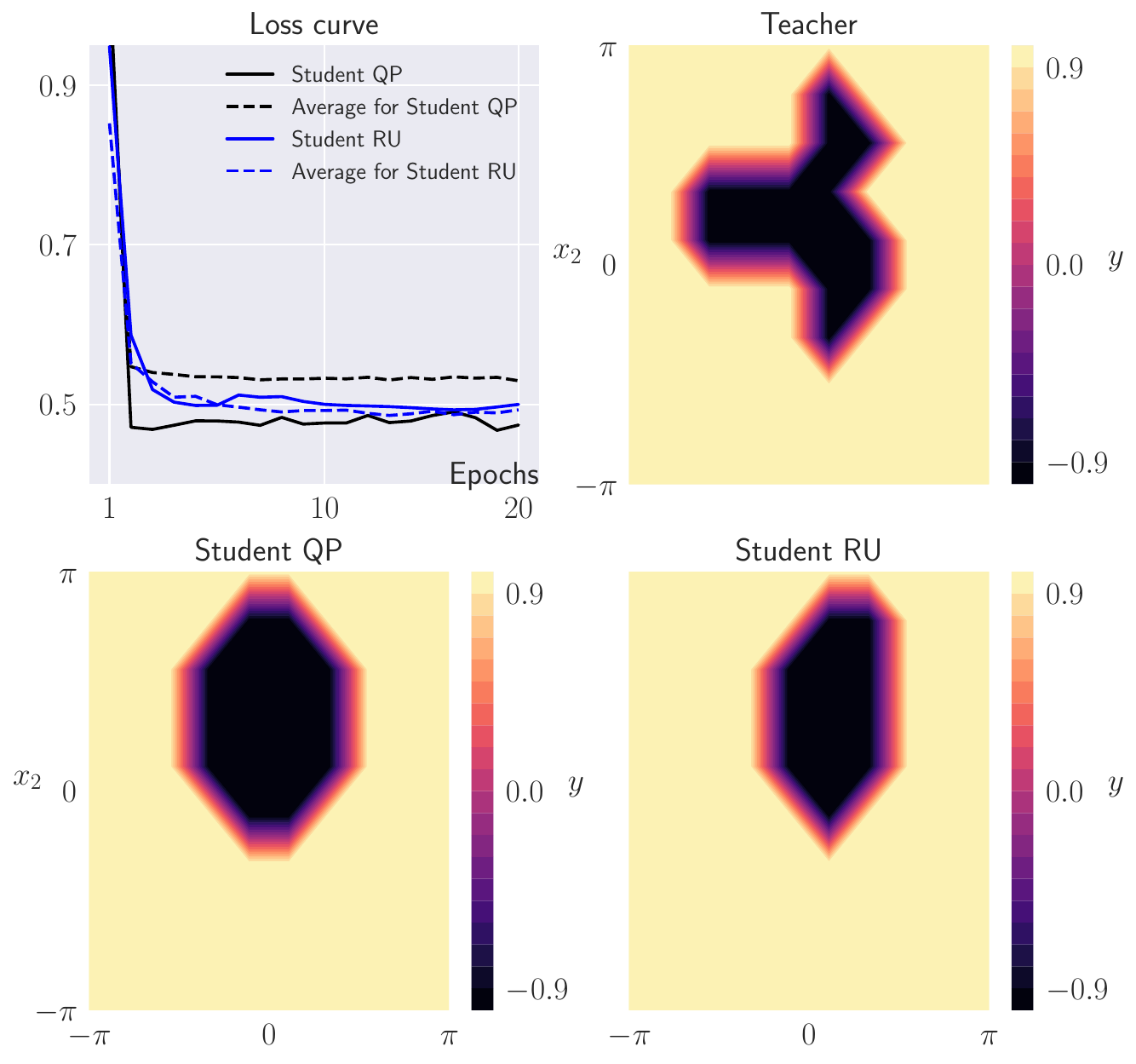}
\caption{The binary prediction maps of the teacher with the deep architecture in Fig.~\ref{fig3} and the students: Student - QP and Student - RU as in Fig.~\ref{fig1}. In the upper right corner, we plot the accuracy curves for the QP (Student - QP black line), RU (Student - Re-uploading blue line) and the student averages over 10 different realizations of the teacher (Student - QP black dashed line and Student - Re-uploading blue dashed line). }
\label{fig7A-all}
\end{figure}
\begin{figure}
\includegraphics[scale=1.0]{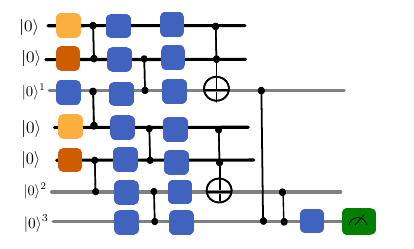}
\caption{An alteration of the QP with many processing and entangling gates. }
\label{random-QP1} 
\end{figure}

In Fig.~\ref{fig2A-all} and \ref{fig7A-all} we present the prediction maps and loss curves with binary labels of the teachers for the students of the toy model (Fig.~\ref{fig1}) and the deep-shallow architectures (Fig.~\ref{fig3}), respectively.
In Fig.~\ref{random-QP1}, we present the alteration of the QP with many processing and entangling gates. This alteration does not improve the performance of the model. This again validates that the functions that can be learned are definitely determined by the number of times the data are encoded in the circuit~\cite{Schuld_Kiloran} and not the amount of deferred measurements.
\subsection{Labelling matters}\label{Labelling}
\begin{figure}
\includegraphics[scale=0.38]{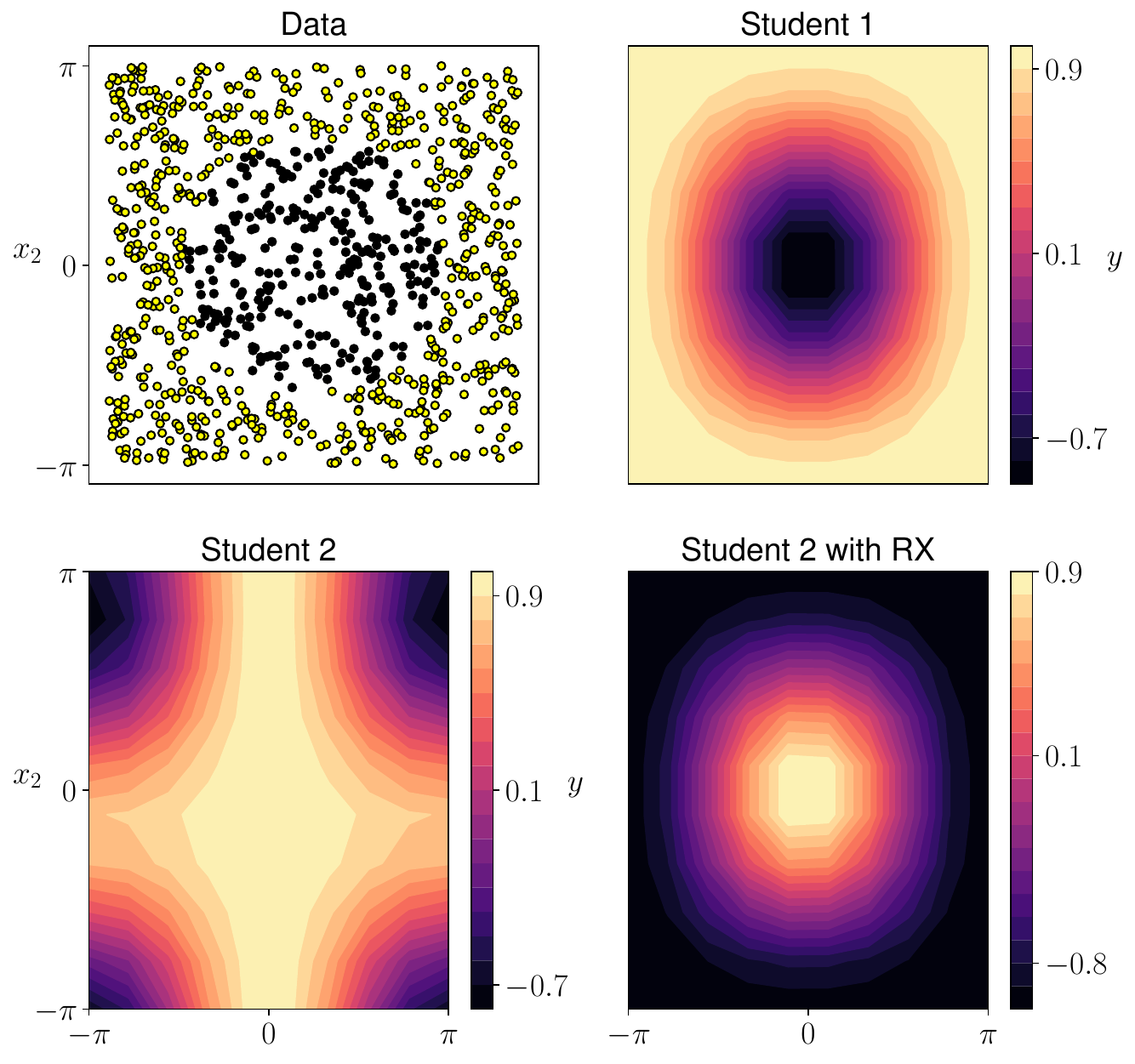}
\caption{We present the input data along with the prediction maps for QP students with different labelling, i.e. Student 1 has labelling -1 for the inner (black) and 1 for the outer circles (yellow), while Student 2 has exactly the opposite. Student 2 with RX has the same labelling as Student 2, but a RX gate is applied before the measurement.}
\label{fig11}
\end{figure}
In Fig.~\ref{fig11}, we show how the labelling affects the results. We plot the prediction map for the circular data with labels $-1$ for the inner (black) and $1$ for the outer circles (yellow) (Student 1 in Fig.~\ref{fig11}). Then, we plot the prediction map for the student with opposite labelling of the data,  i.e. $1$ for the inner (black) and $-1$ (yellow) for the outer circles (Student 2 in Fig.~\ref{fig11}). As shown, the prediction map of Student 2 with RX is very different from the circular structure of the data and the training is not successful. 

The final state of the ancilla qubit is of the form $\ket{\psi} = \alpha \ket{0} + \beta\ket{1}$ and the measurement gives $\langle Z \rangle = \alpha^2 - \beta^2$. Therefore, for the label $-1$, $\langle Z \rangle$ should be close to $-1$. The average probability vector of the ancilla qubit for the label $-1$ is $P_{-1} = [\alpha^2, \beta^2]= [0.37, 0.64]$, and indeed, $\alpha^2 - \beta^2<0$. Then, the training is successful and the prediction map of Student 1 is similar to the data structure. If we change the labelling, we get the following average probability vector $P_{-1} =  [0.94, 0.06]$ for the label $-1$, where now $\alpha^2 - \beta^2>0$. This is the reason why the training fails for Student 2, which is fixed by adding a PauliX gate before the measurement (see Student - Re-uploading with RX in Fig.~\ref{fig11}).

\subsection{Input data}\label{input_data}
\begin{figure}
\includegraphics[scale=0.38]{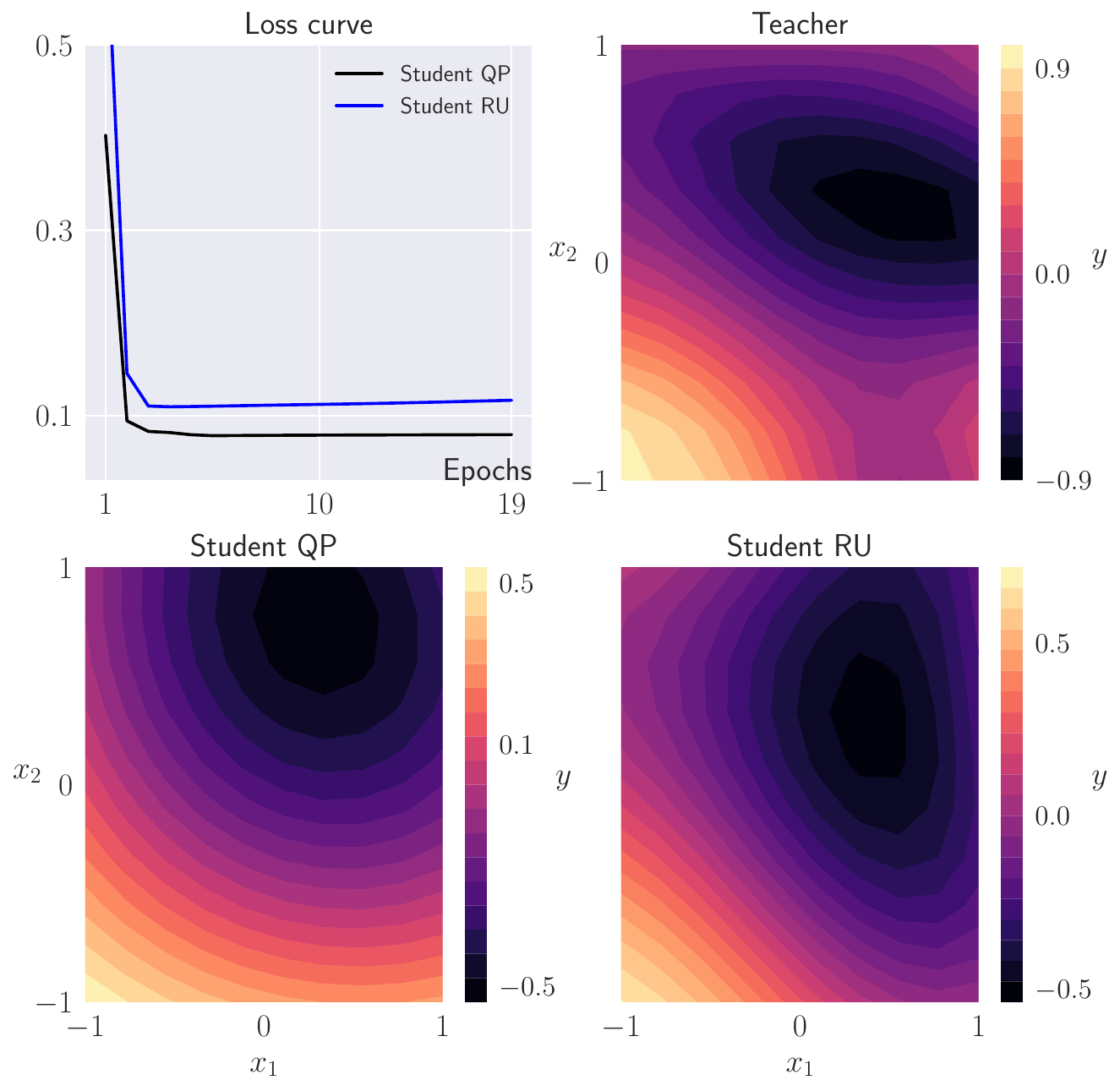}
\caption{The prediction maps of the teacher with the deep architecture in Fig.~\ref{fig3} and the students: Student - QP and Student - RU as in Fig.~\ref{fig1} but with normalization $[-1,1]$ instead of $[-\pi, \pi]$. In the upper right corner, we plot the loss curves for the QP (Student - QP black line) and the RU (Student - Re-uploading blue line).}
\label{fig5-zoomin}
\end{figure}
As emphasized in the main text, the normalization of the input data highly affect the training and success of the circuits, since the quantum functions are periodic in the input data. For example, we train the students in Fig.~\ref{fig3}, but with a different normalization, i.e. instead of $[-\pi, \pi]$ with $[-1,1]$. As it can be seen in Fig.~\ref{fig5-zoomin}, the prediction maps do not show the whole structure of the data and both students seem to perform well, contrary to Fig.~\ref{fig3} where Student - Re-uploading performs better. Therefore, one should be careful when choosing the normalization of the input data.

\end{document}